\documentclass[%
showpacs,
 amsmath,amssymb,
 aps,
pra,
]{revtex4-1}

\usepackage{graphicx}
\usepackage{dcolumn}
\usepackage{bm}
\usepackage{hyperref}

\begin{document}

\preprint{APS/123-QED}

\title{Elliptic solitons in optical fiber media}

\author{D\'efi Jr. Jubgang Fandio}
\email{defandio@yahoo.fr}
\author{Alain M. Dikand\'e}
\email{dikande.alain@ubuea.cm}
\affiliation{Laboratory of Research on Advanced Materials and Nonlinear Sciences, Department of Physics, Faculty of Science, University of Buea P.O. Box 63, Buea Cameroon}%
\author{A. Sunda-Meya}
\email{asundame@xula.edu}
\affiliation{Department of Physics, Xavier University of Louisiana, 1 Drexel Drive New Orleans, LA 70125}%

\date{\today}

\begin{abstract}We examine the evolution of a time-varying perturbation signal pumped into a mono-mode fiber in the anomalous dispersion regime. We analytically establish that the perturbation evolves into a conservative pattern of periodic pulses which structures and profiles share close similarity with the so-called soliton-crystal states recently observed in fiber media [see e.g. A. Haboucha et al., Phys. Rev. A\textbf{78}, 043806 (2008); D. Y. Tang et al., Phys. Rev. Lett. \textbf{101}, 153904 (2008); F. Amrani et al., Opt. Express \textbf{19}, 13134 (2011)]. We derive mathematically and generate numerically a crystal of solitons using time division multiplexing of identical pulses. We suggest that at very fast pumping rates, the pulse signals overlap and create an unstable signal that is modulated by the fiber nonlinearity to become a periodic lattice of pulse solitons which can be described by elliptic functions. We carry out a linear stability analysis of the soliton-crystal structure and establish that the correlation of centers of mass of interacting pulses broadens their internal-mode spectrum, some modes of which are mutually degenerate. While it has long been known that high-intensity periodic pulse trains in optical fibers are generated from the phenomenon of modulational instability of continuous waves, the present study provides evidence that they can also be generated via temporal multiplexing of an infinitely large number of equal-intensity single pulses to give rise to stable elliptic solitons. 
\end{abstract}

\pacs{42.55.Wd, 42.65.Tg, 42.70.Mp}
\maketitle

\section{\label{sec:level1} Introduction}
Optical sources able to generate wavelength tunable ultra-short pulse trains with high quality and high-repetition rates have attracted a great deal of attention in the recent past \cite{agr1,agr2}. These specific structures find important application in optical networks combining wavelength division multiplexing and optical time domain multiplexing transmission techniques \cite{agr1,agr2}. Traditionally soliton operations in optical media have been associated with a net negative dispersion, given that the balance between a negative dispersion and the nonlinear optical responses (Kerr and non-Kerr) of the media is more likely to promote nonlinear transform-limited pulses \cite{mol}. \\
Although a full description of most real optical fiber networks as well as fiber laser media generally requires the account of several complex processes interfering in the system dynamics (high-order dispersions processes, loss and gain dynamics and so on), the fundamental optical soliton (pulse-shape signal) observed in these systems is generic from the nonlinear Schr\"odinger equation \cite{menyuk}. Among other possible optical soliton structures that can also be derived from the nonlinear Schr\"odinger equation multi-soliton complexes are particular, due to their suitability for high-bit and high-repetition rate data transfer technology. These include harmonically mode-locked vector solitons, soliton crystals and bound soliton states representing a periodic lattice of pulses distributed in a one-dimensional chain and that have recently been the subject of intense theoretical investigations, in connection with a broad range of optical-wave transmission phenomena \cite{tang,kart1,kart2,kart3,dik1,dik2,zhao}. \\ 
Our main point of focus in this work are the periodic trains of high-intensity signals~\cite{amra1,amra2,abouch1,abouch2,abouch3} that have been reported recently in passively mode-locked double cladd fiber lasers, where each pulse in the trains appears to be a double time scale comb-like signal consisting of a pulse soliton hosting a quasi-periodic pattern of solitons at its apex and whose self-ordering may stem from the phase-locking of wave components modulated by gain variation~\cite{abouch1,abouch2}. To this last point theoretical investigations of the stability of steady wave spread in optical fibers have long predicted~\cite{agr1,agr2} that the nonlinear Schr\"odinger equation could admit a family of periodic soliton solutions represented by Jacobian elliptic functions. On the other hand it is well established that in the anomalous regime of dispersion, modulational instability can generate a web of bound pulses whose inter-pulse length is controlled by an external mechanism \cite{chern}. In general high-power input pulses are necessary to trigger pulse interaction but the effect is relatively unstable and the bound pulse train eventually splits into independent pulses after few kilometers of propagation. Methods to compensate for this inconvenience might be enclosing the fiber into a Fabry-Perot resonator, or by using dispersion-managed systems. The gain acquired by modulational instability in the first method or by timing-jitter reduction, as illustrated in \cite{mu} in the second case, provides enough energy for pulse interaction. Pulse repetition control can be achieved by means of self-starting laser pumps \cite{menyuk1,chen} able to generate ultra-short high energetic pulses into the fiber.\\
In this paper we examine the intimate structure (width, tails and mutual separation between pulses) of the periodic trains of pulses reported recently in fiber lasers and called soliton-crystal signals~\cite{amra1,amra2,abouch1,abouch2,abouch3}. In particular we wish to establish unambiguously the connection between the elliptic-wave solution of the cubic nonlinear Schr\"odinger equation, and the pattern formed by means of time division multiplexing of pumped pulses \cite{menyuk} thus emphasizing the minor role of the gain dynamics in the formation mechanism of soliton crystals while the gain is expected to hold a more major role in the pulse interaction stengths and the soliton-crystal stability in general. \\ 
In section \ref{Sec:level2} we present the model consisting of a cubic nonlinear Schr\"odinger equation for a single-mode fiber, and find exact nonlinear periodic-wave solution representing elliptic soliton structures. In section \ref{Sec:level3} we revisit a familiar reconstruction scheme (see e.g. \cite{menyuk,abouch3}) in order to identify the pulse distribution in the soliton crystal. In this respect we consider a standard ansatz \cite{menyuk,abouch3} describing a multiplex state of periodically spaced solitons, that can be obtained by repetitively pumping an infinitely large number of pulses into the fiber. We expand the ansatz and find an analytical wave function that permits a direct comparision with the elliptic-soliton solution of the nonlinear propagation equation. Next a numerical implementation of the time division multiplexing of identical pulses is carried out now with a finite number of pulses, to illustrate the consistency of the proposed theoretical reconstruction scheme with experimental results~\cite{amra1,amra2}. In section \ref{Sec:level4} we carry out a linear stability analysis to check the robustness of the soliton-crystal structure against small-amplitude noises, in this context we point out the implication of the multiplexing of equal-intensity pulses on the stability of individual pulses as well as the multiplex state. Concluding remarks follow in section \ref{Sec:level5}.

\section{\label{Sec:level2} The model and elliptic-soliton structures}
\begin{figure*}
 \begin{minipage}[c]{0.51\textwidth}
\includegraphics[width=3.in,height=2.in]{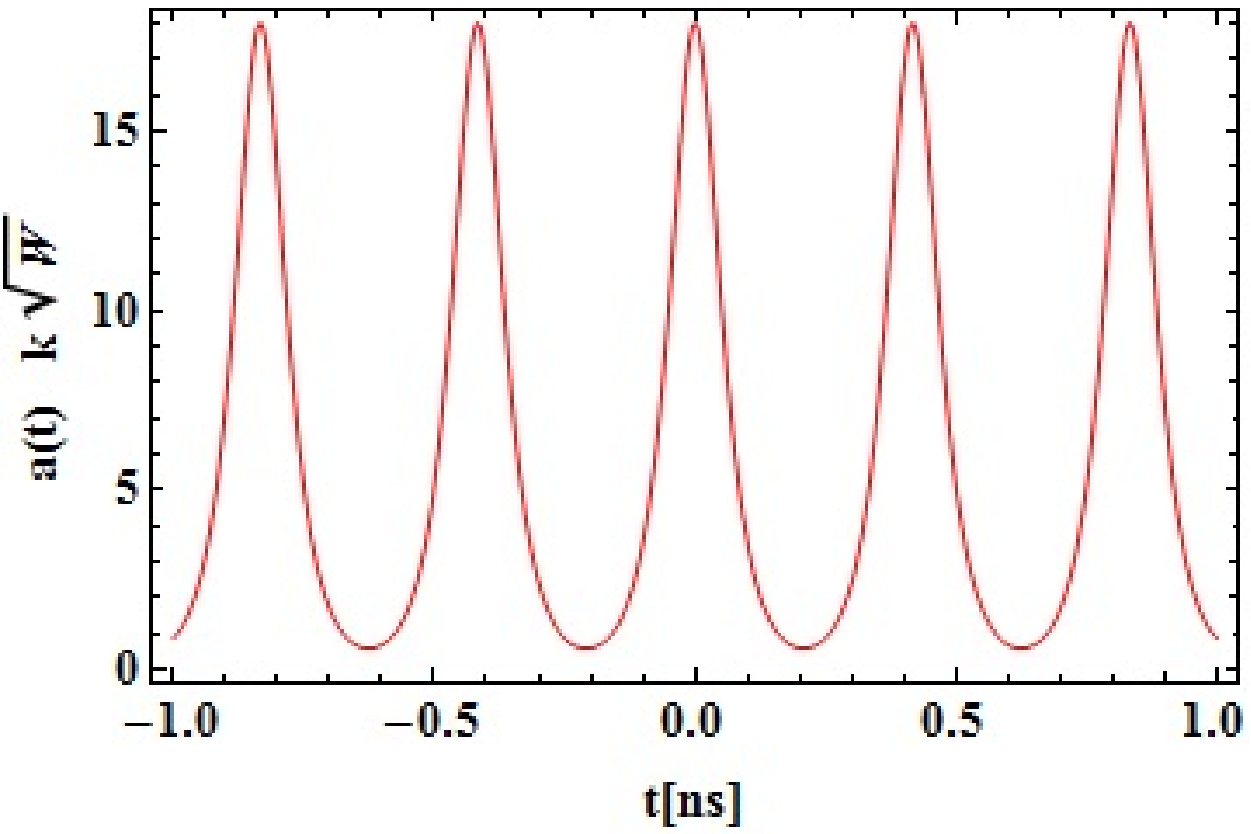}(a)
\end{minipage}%
\begin{minipage}[c]{0.51\textwidth}
\includegraphics[width=3.in,height=2.in]{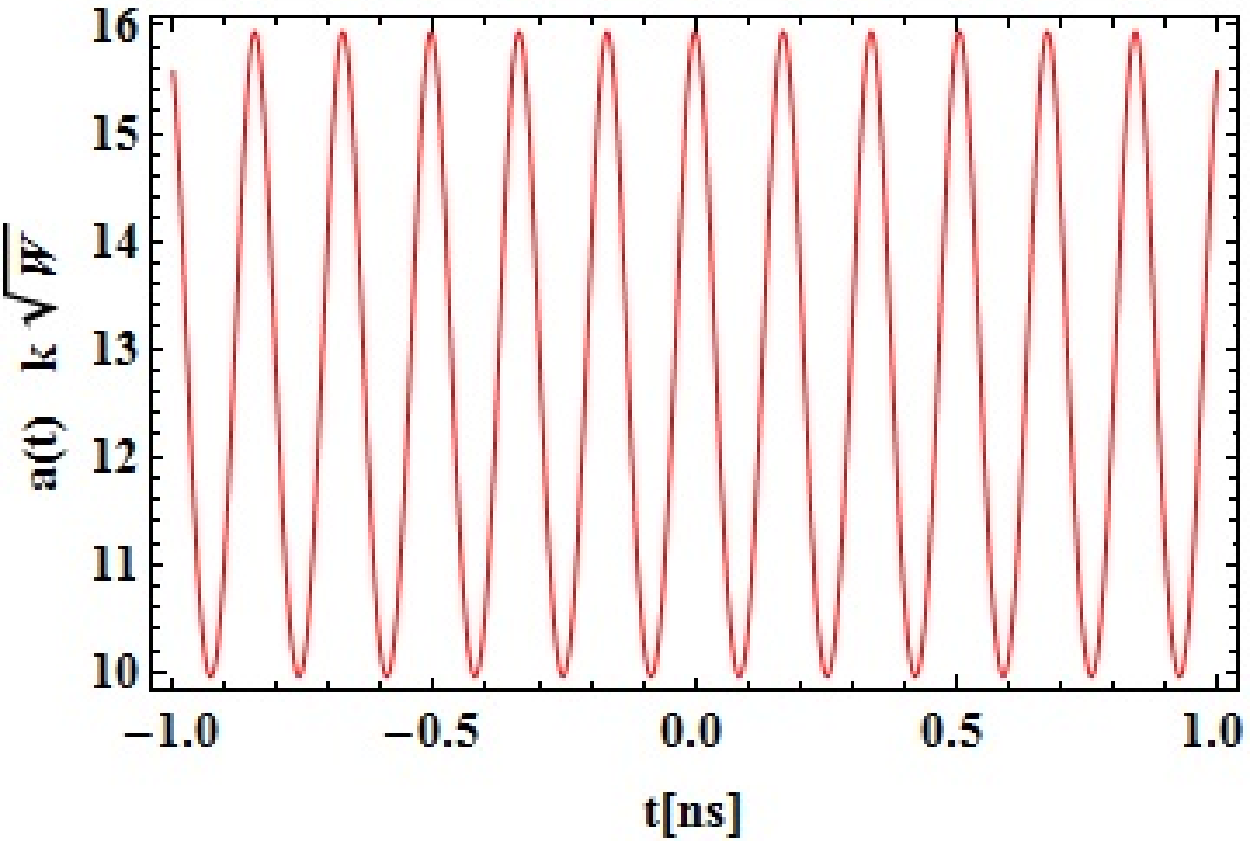}(b)
\end{minipage}
\caption{\label{fig:fandio1} Elliptic-soliton solution to the nonlinear amplitude equation (\ref{eq:zero}), for two different values of the parameter of pulse dispersion $k$: (a) Periodic network of sharp pulse signals for large dispersion parameter ($k=0.98$), (b) Elliptic-wave decay into a quasi-continuous signal for small dispersion parameter ($k=0.6$).}
\end{figure*}

The propagation of slowly varying wave envelopes in optical fibers exhibiting weak dispersion and weak nonlinearity, is governed by the cubic nonlinear Schr\"odinger equation:
 \begin{equation}
  i\frac{\partial q(z,t)}{\partial z} - \frac{1}{2}\beta_2 \frac{\partial^2 q(z,t)}{\partial t^2} + \gamma|q(z,t)|^2q(z,t) = 0,\label{eq:one}
  \end{equation}
where $z$ is the spatial coordinate of the envelope, $t$ is the propagation time, $\beta_2$ is the group velocity dispersion of the fiber material and $\gamma$ is the nonlinear Kerr coefficient accounting for self-phase modulation. The slowly varying wave envelope $q(z,t)$ is normalized such that the square of its magnitude represents the power transmitted throughout the fiber. For weak fields, nonlinear effects are neglected and the solution to eq(\ref{eq:one}) is a steady-state wave that disperses along the fiber core and vanishes in the cladding. \\ We assume that the slowly varying amplitude $q(z,t)$ injected by the pump is a strong time-varying perturbation that temporally awakes nonlinear effects in the fiber, in this respect we set $q(z,t) =  a(t)e^{i\beta z}$ where $\beta$ is the wave parameter. The amplitude $a(t)$ of the signal governing the temporal evolution of the envelope then obeys the equation:
\begin{equation}
\left(\frac{da}{dt}\right)^2= -\frac{2\beta}{\beta_2}a^2 + \frac{\gamma}{\beta_2}a^4 + C,\label{eq:zero}
\end{equation}
where $C$ is the constant of energy integral that determines profiles of the amplitude $a(t)$. For a localized profile we expect a rapid evanescence of the wave outside its time-bandwidth such that the constant $C$ tends to zero. In the anomalous dispersion regime ($\beta_2<0$), the localized-wave solution is a single hyperbolic-secant pulse given by:
\begin{equation}
a(t)=  \sqrt{\frac{2\beta}{\gamma}}\ sech\left[\sqrt{\frac{-2\beta}{\beta_2}}(t-t_0)\right], \label{eq:0}
\end{equation}
the shape of which is independent on the autocorrelation trace and propagation distance such that its "soliton" identity is always preserved. When the constant $C$ is non-zero and negative, energetic conditions become detrimental to the hyperbolic-secant pulse. However we can still find nonlinear solutions to the amplitude equation as:
\begin{equation}
 a(t) = \sqrt{\frac{2\beta}{\gamma \ (2-k^2)}}\, dn\left[\sqrt{\frac{-2\beta}{\beta_2(2-k^2)}}(t-t_0), k \right]. \label{eq:two}
 \end{equation}
In the last formula $dn$ is the Jacobi elliptic delta function of modulus $k$ ($0<k\leq 1$), while $t_0$ is an arbitrary initial time. \\
The Jacobi elliptic $dn$ function is periodic in its time argument $t$ with a time period:
\begin{equation}
\tau = 2K(k)\sqrt{\frac{\beta_2(2-k^2)}{-2\beta}}, \label{eq:three}
\end{equation}
with $K(k)$ the Elliptic Integral of first kind. In fig. \ref{fig:fandio1}, we sketched the autocorrelation trace of the Jacobi elliptic delta function signal (\ref{eq:two}). 
Possible numerical values for characteristic parameters are $\beta=1.55\mu m$, $\beta_2 = -15ps^2/km$ and $\gamma = 25 W^{-1}/km$ and correspond to typical data common in most fiber laser setups~\cite{,amra1,amra2,abouch1,abouch2}. As fig. \ref{fig:fandio1} shows, the $dn$ function describes bound states of periodically spaced pulses forming a \textit{soliton crystal} \cite{kart1,kart2,kart3,dik1,dik2}. Remarkably small values of the modulus $k$ tend to increase the pulse density in the crystal whereas large values of this parameter isolate each pulse, as a matter of fact when $k=1$ the soliton crystal breaks down and the single-pulse regime is favored. Its turns out that the modulus $k$ can be readily regarded as the parameter of pulse dispersion. \\
The presence of the autocorrelation propagation parameter of the envelope, as well as its conservative nature manifested in the distance-independent power transmitted, are relevant characteristics providing the envelope eq. (\ref{eq:two}) with a "soliton" feature.  We can therefore conclude that the Jacobi $dn$ function stands for a good mathematical representation of a soliton crystal. Since the generic wave equation (\ref{eq:one}) is integrable, all physical features arising from the propagation of waves will be intrinsic to such system such that the soliton crystal structure is clearly an essential profile of light envelopes in Kerr nonlinear optical media.

\section{\label{Sec:level3} Soliton-crystal reconstruction}
\subsection{\label{subsec:level3a}Analytical reconstruction}
\begin{figure}
 \includegraphics[width=3in, height=2.0in]{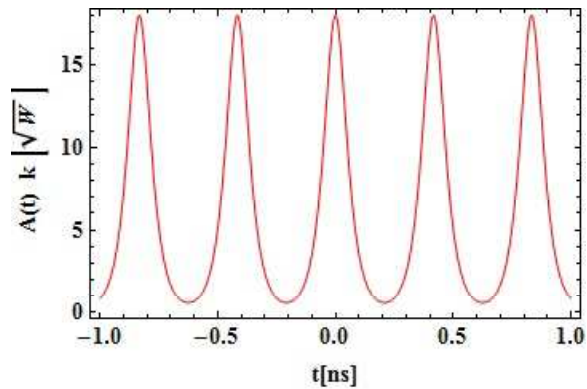}
 \caption{\label{fig:fandio2}Autocorrelation trace of the periodic train of bound solitons, obtained from analytical reconstruction by time division multiplexing of identical pulses (eq. \ref{eq:five}).}
\end{figure}

It has long been established~\cite{agr2,chern,mu} that nonlinear-wave generation in optical fibers results from the modulational instability of continuous waves. The modulational instability in this case can lead either to a single pulse \cite{menyuk1}, as for instance a strong hyperbolic secant pulse produced by the superposition of continuous waves oscillating at a constant frequency shift, or to a train of pulses ~\cite{milon,akhmed}. It would however be interesting to examine the structure of a periodic train of pulses constructed by time division multiplexing~\cite{menyuk} of an infinitely large number of pulses, which in fact would be the output signal resulting from the interaction of pulses pumped into the fiber by a pulse-mode laser at a given period. In this last goal we consider a laser pump injecting pulses into a standard mono-mode fiber at finite wavelengths, in the weak anomalous dispersion regime $\beta_2$. Let the hypothetical time multiplexed output signal A(t) be expressed as ~\cite{menyuk,abouch3}:
\begin{equation}
A(t) = \sum\limits_{n=-\infty}^{\infty} \sqrt{\frac{2\beta_m}{\gamma}}\, sech\left[\sqrt{\frac{-2\beta_m}{\beta_2}}(t-t_0-n\tau_A) \right],\label{eq:four}
\end{equation}
 where the subscript $m$ refers to the propagation mode and $\tau_A$ is the period at which the pulses are pumped into the fiber. In the case of multi-mode transmission the sum in eq. (\ref{eq:four}) is not integrable \cite{dik1} and thus does not yield an exact solution. However if we consider the case of identical pulses ($\beta_m=\beta$) this sum becomes exact forming the bound states:
\begin{equation}
A(t) = 2\sqrt{\frac{\beta_2}{\gamma}} \frac{K(k)}{\tau_A}\, dn\left[ 2 K(k)\frac{t}{\tau_A}, k\right]. \label{eq:five}
\end{equation}
According to eq (\ref{eq:five}) as well as fig.\ref{fig:fandio2}, pulses periodically pumped into the optical fiber interact instantaneously to form a periodic pattern of identical pulse solitons under the anormalous regime of dispersion. 
Because of the similarity between formula (\ref{eq:five}) and the elliptic-wave solution to the nonlinear amplitude equation obtained in the previous section, we can state unambiguously that the soliton-crystal state created by time division multiplexing has the same structural feature as the elliptic-wave solution to the propagation equation. Namely we also expect the time division multiplexed pulses state to decay into a quasi-continuous wave signal when the dispersion parameter $k$ decreases from its maximum $k=1$, while pulses in the soliton crystal will be sharp and well separated as $k$ tends to one. It is relevant at this step to stress that from the standpoint of exact mathematical treatment the sum of an infinite number of secant hyperbolic pulses as given by formula (\ref{eq:four}), to obtain the elliptic-wave structure formula (\ref{eq:five}), must be supplemented with the existence condition:
\begin{equation}
\tau_A=\frac{\pi K(k')}{\lambda K(k)}, \label{cond} 
\end{equation}
where
\begin{equation}
\lambda=\sqrt{\frac{-2\beta}{\beta_2}}, \quad \hskip 0.1truecm k'=\sqrt{1-k^2}, \label{eq:fivea}
\end{equation}
that links the pumping period $\tau_A$ to the width at half height of individual pulses $\lambda$ in the pulse multiplex. This existence condition also determines the stability of the soliton-crystal state, indeed when $k$ tends to  one $\tau_A$ tends to infinity which is consistent with the fact that $dn$ tends to $sech$ when $k \rightarrow 1$. On the other hand, as $k$ decreases $\tau_A$ also decreases such that the separation between pulses is not large enough for the pulses to preserve their full identity in the soliton-crytal state, causing a decay of the soliton crystal into quasi-continuous wave signals.\\ 
Experimental evidences on soliton-crystal structures \cite{abouch1,abouch2,abouch3,amra1,amra2} have established that high pumping rates favor solitons interaction and increase the intensity of the soliton crystal. When the group velocity dispersion is sufficiently high, the input pulses overlap and become a bound state that is modulated by the fiber nonlinearity into a chain of periodic pulses \cite{liu}. It is remarkable from formula (\ref{eq:five}) that the amplitude of the time division multiplexed signal is inversely proportional to the puming period $\tau_A$ which depends solely upon the pump. In particular this formula suggests that to generate a regular well-shaped pulse train practically it is desirable to pump high energetic pulses with the shortest widths possible, and passively mode-locked lasers are suitable for such pump because they can repetitively generate pulses of the order of femtosecond~\cite{menyuk1,chen}. Titanium-doped sapphire lasers passively mode-locked by use of slow saturable absorbers, for instance, are able to generate 10 to 30 fs pulses. Furthermore self-starting passively mode-locked lasers that continuously generate trains of pulses once turned on appear to be good candidates for pumping periodically a large number of pulses in the process of time division multiplexing \cite{menyuk,chen}, in this last context the separation between pulses would theoretically correspond to the round trip time of the laser cavity as observed in the harmonic mode-locking-induced periodic bunching of pulse signals~\cite{amra1,niang1}.
\subsection{\label{subsec:level3b}Numerical reconstruction}
\begin{figure*}
 
\begin{minipage}[c]{0.51\textwidth}
\includegraphics[width=3in,height=2.0in]{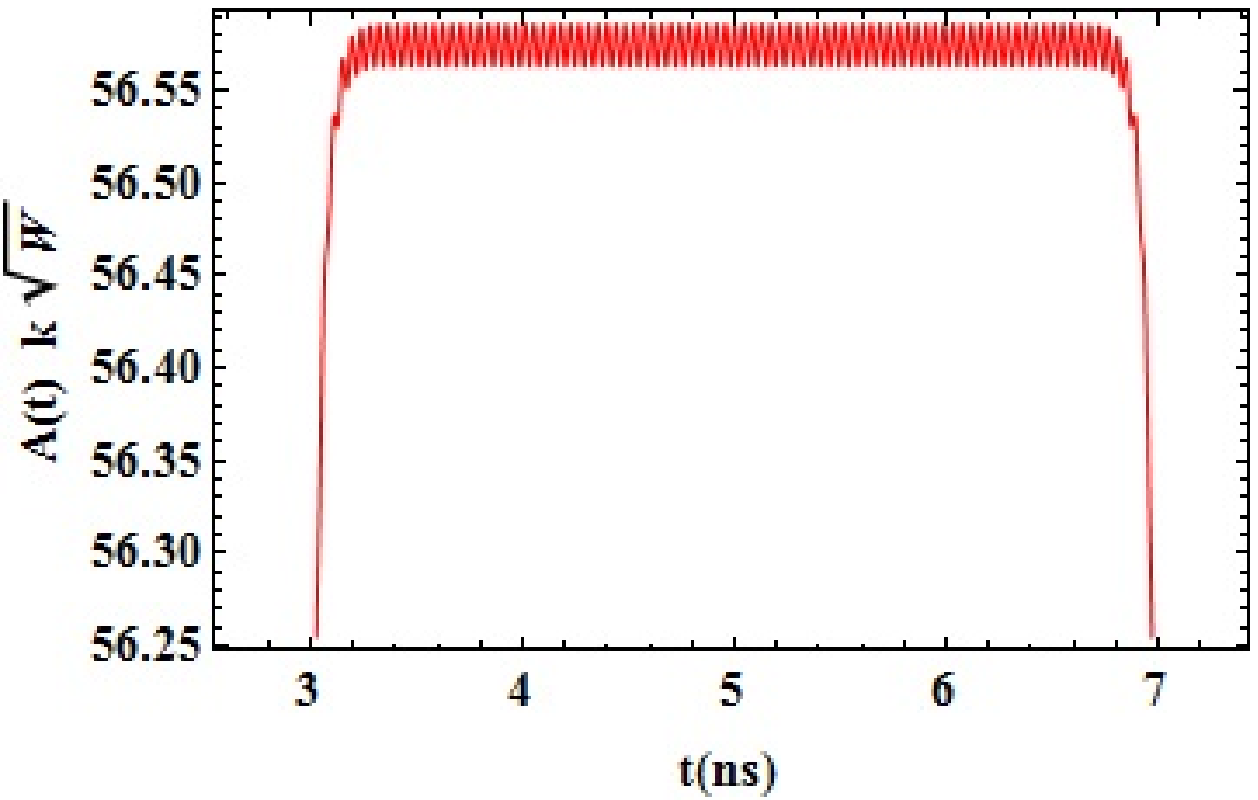}(a)
\end{minipage}%
\begin{minipage}[c]{0.51\textwidth}
\includegraphics[width=3in,height=2.0in]{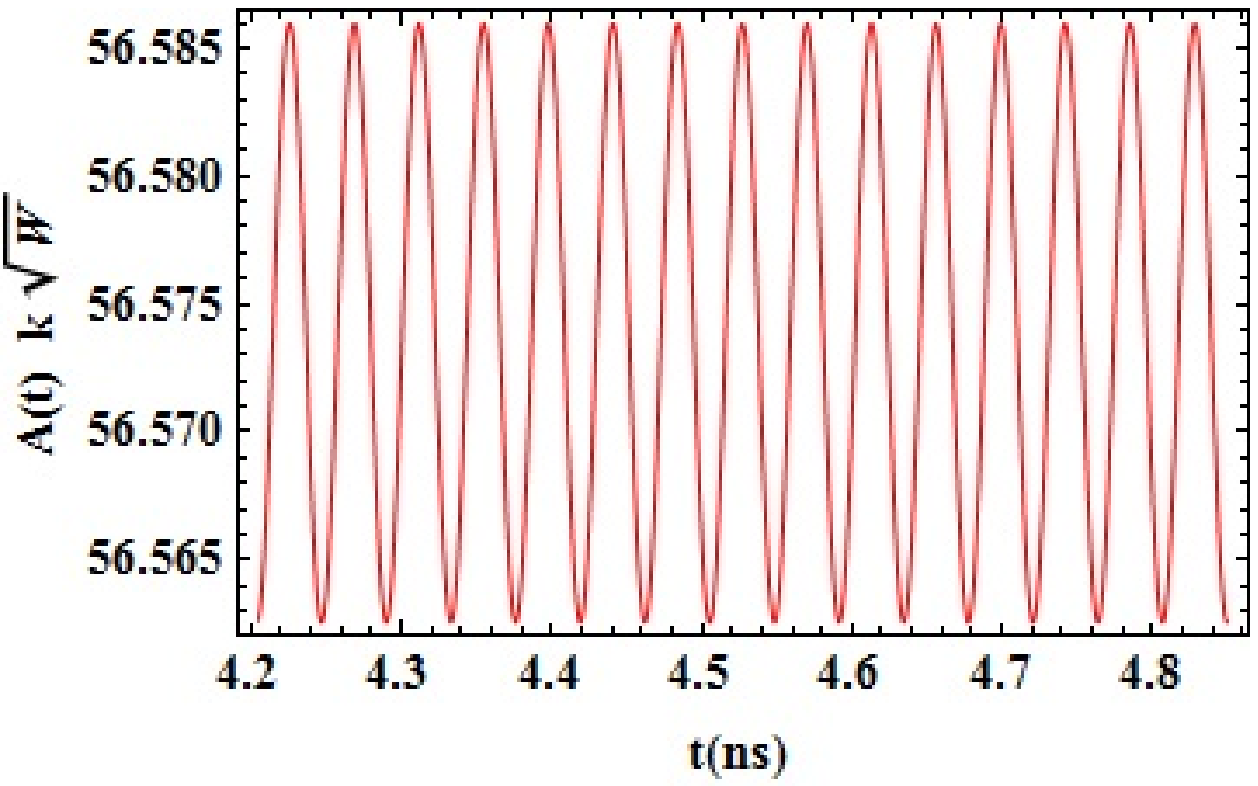}(b)
\end{minipage}
\caption{\label{fig:fandio3} Numerically reconstructed signal at the pumping rate $\tau_A= 23.248 \ 1/ns$: (a) Double time scale comb-like pulse, (b) Apex of the large pulse: Soliton crystal of finite length.}
\end{figure*}

\begin{figure*}
\begin{minipage}[c]{0.51\textwidth}
\includegraphics[width=3in,height=2.0in]{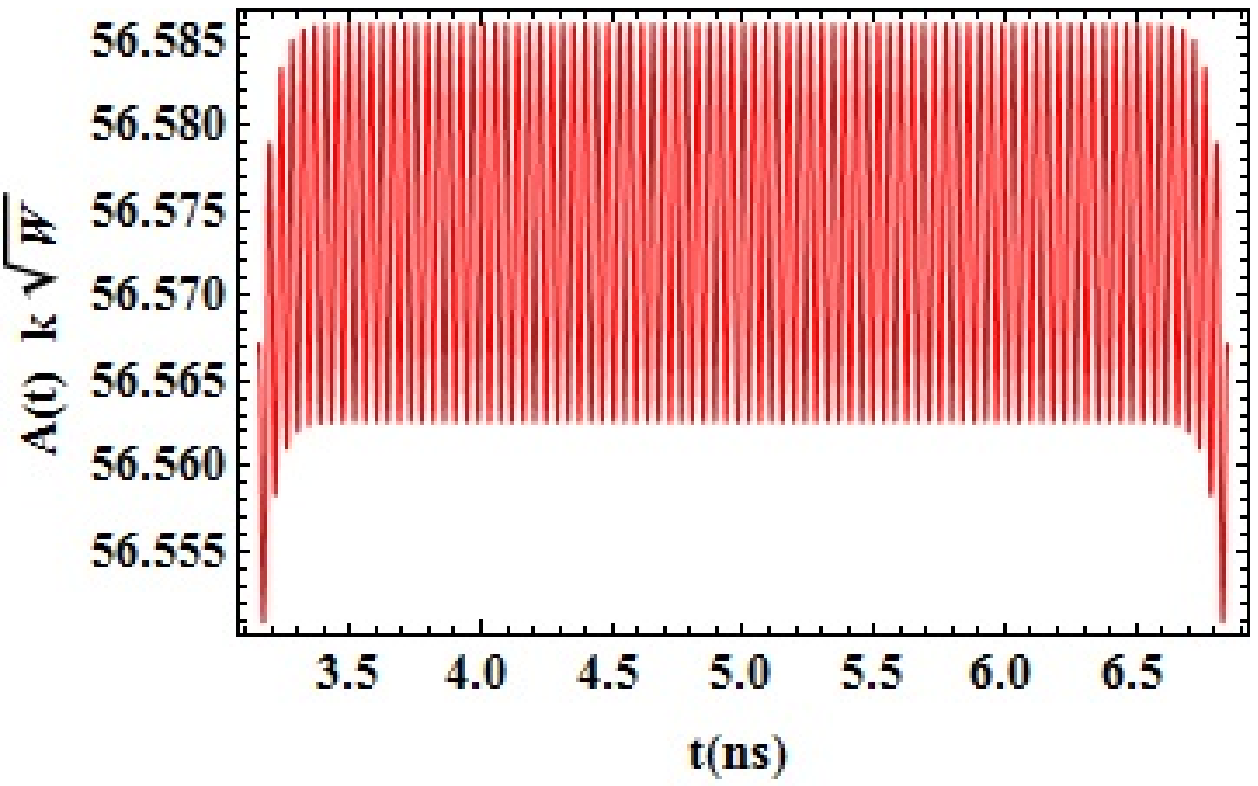}(a)
\end{minipage}%
\begin{minipage}[c]{0.51\textwidth}
\includegraphics[width=3in,height=2.0in]{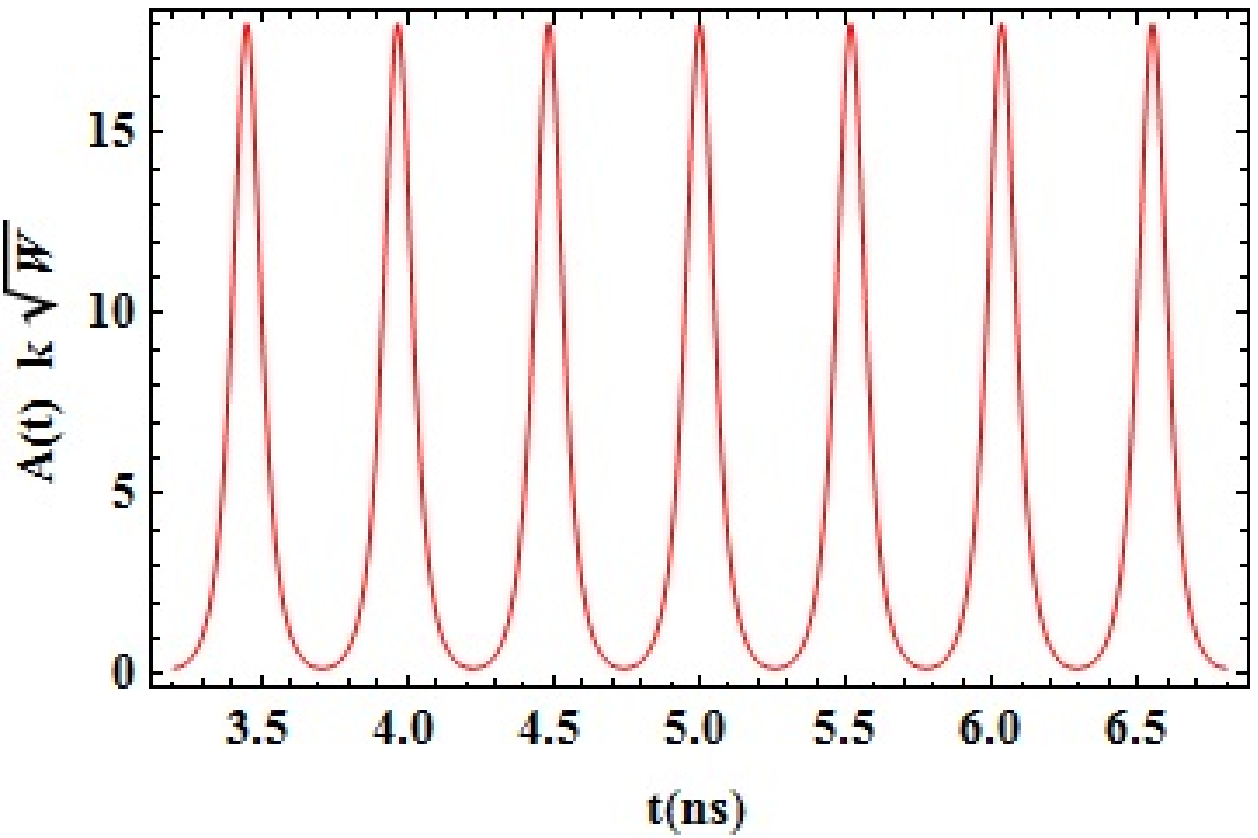}(b)
\end{minipage}
\caption{\label{fig:fandio4}Autocorrelation trace of the reconstructed signal: (a) Strong modulation illustrated by the amplitude gap between the soliton crystal and the large pulse, (b) Minimal modulation at the threshold frequency.}
\end{figure*}

In experiments on multi-pulse processing it can be necessary to reconstruct the observed periodic pattern of pulses analytically as well as numerically, for a better understanding of their fundamental properties. Indeed, unlike the analytical reconstruction scheme where we needed to sum an infinite number of pulses to be able to find a single function comparable to the elliptic-wave solution to the propagation equation, in the numerical reconstruction we can consider a large but finite number of pulses and comparing the structure obtained from this numerical sum over a finite number of pulses with the elliptic-wave signal, is also relevant. \\
For the numerical reconstruction we use the ansatz:
\begin{equation}
A(t) = \sum\limits_{n=-N/2}^{N/2} \sqrt{\frac{2\beta_m}{\gamma}} sech\left[\sqrt{\frac{-2\beta_m}{\beta_2}}(t-t_0-n\tau_A) \right],\label{eq:six}
\end{equation}
where N is the number of pumped pulses. By considering a hundred of identical pulses ($\beta_m=\beta$)  a two-time-scale signal is produced as seen in fig. \ref{fig:fandio3}: a slow-time scale large pulse, and a fast-time scale soliton crystal. Suppose each pulse in the crystal is $44ps$ of width and is separated from its neighbour by $\Delta t = 40fs$. Evaluating the number of pulses in the crystal whose total length is $5ns$, taking into consideration the pulse width and pulse seperation, yields 113 pulses in the soliton crystal. 
The excess in the number of pulses comes from the non-linear gain induced by the modulational instability of the interacting input pulses. The pulses are not as sharp as those experimentally observed in rare earth doped fiber lasers \cite{abouch1,abouch2,abouch3}, because in the later context the immediate response of the medium to the laser gain dynamics modifies the propagation and the interaction of pulses in the crystal. This is not the case for our optical fiber model where we neglected the laser gain dynamics, so pulses are less intense and the autocorrelation compels them to a broader profile. In fact the formation of the analytically obtained time-division multiplexed signal, confirmed by the numerical reconstruction, perfectly agrees with the theory that pulse interaction caused by the overlap of pulses at the fiber input end generates an unstable signal that is reformed into a one-dimensional lattice of pulses. The pattern owes its periodic nature to the constant rate at which identical pulses are pumped into the fiber. A close observation of the periodic pattern behaviour, however, reveals a peculiar behaviour: pulse interaction is inversely proportional to the pumping period and we expect it to be minimal when the pumping period falls below the temporal width of each pulse, such that a pulse entering the optical fiber collides with its predecessor. However the numerical reconstruction illustrated by fig. \ref{fig:fandio4} shows a strong modulational instability when the pumping period is of the order of the pumped pulse width. 
The strength of the modulation can be qualitatively evaluated by the amplitude difference between the floor of the giant pulse and the lowest point of the soliton crystal. From a theoritical stand point, pulses start to interact when their tails overlap and we numerically determined that for our system this interaction starts when the separation time between pulses is a dozen times the width of each pulse as shown in fig. \ref{fig:fandio4}(b). This observation is manifest of the wide bandwidth pulse solitons display in optical fibres.
\section{\label{Sec:level4} Linear stability analysis of the Soliton-crystal structure}
\begin{figure*}
\begin{minipage}[c]{0.51\textwidth}
\includegraphics[width=2.5in,height=1.8in]{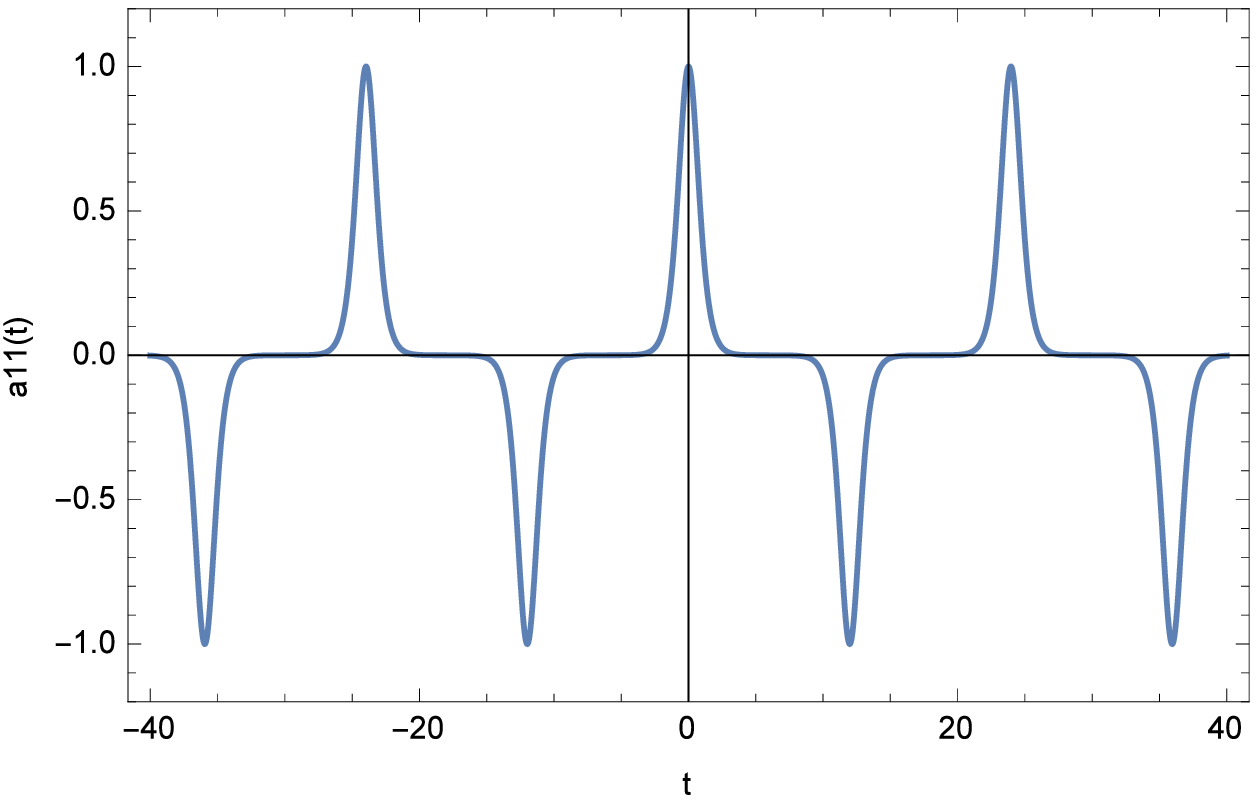}
\end{minipage}%
\begin{minipage}[c]{0.51\textwidth}
\includegraphics[width=2.5in,height=1.8in]{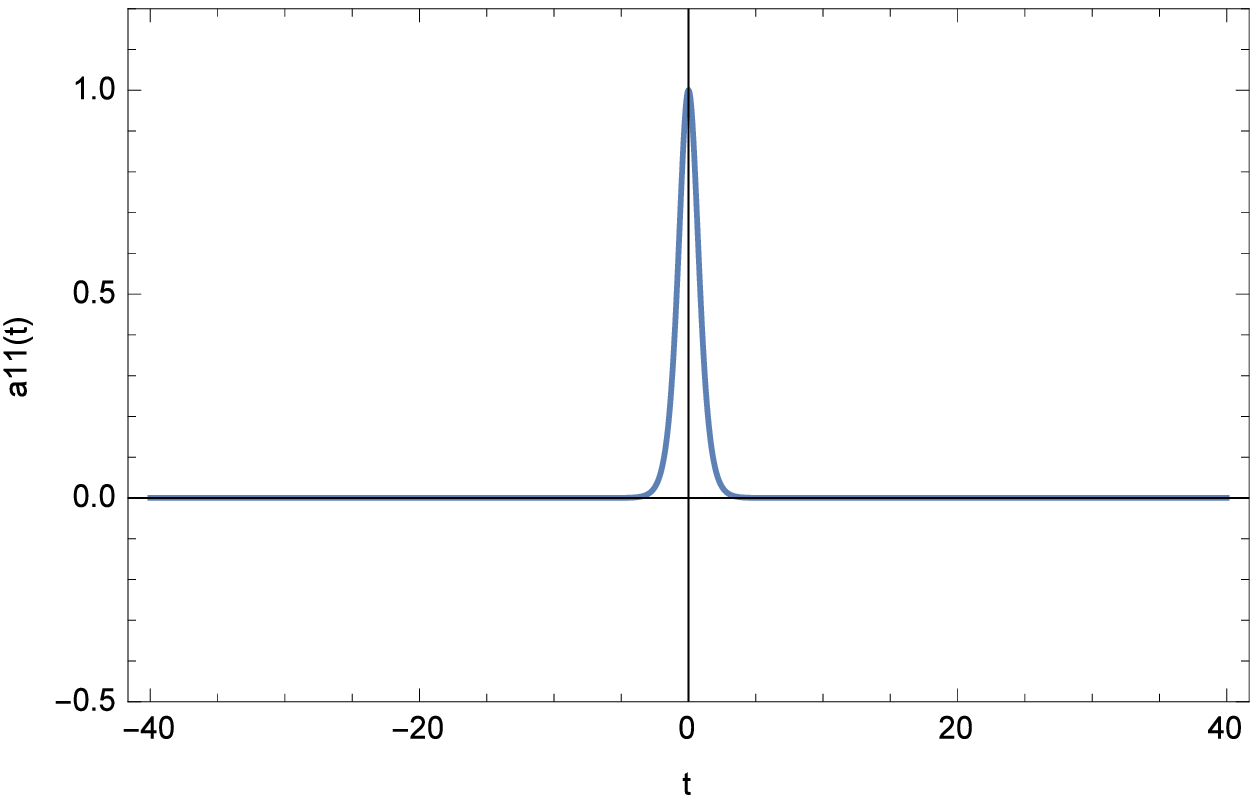}
\end{minipage}\\
\begin{minipage}[c]{0.51\textwidth}
\includegraphics[width=2.5in,height=1.8in]{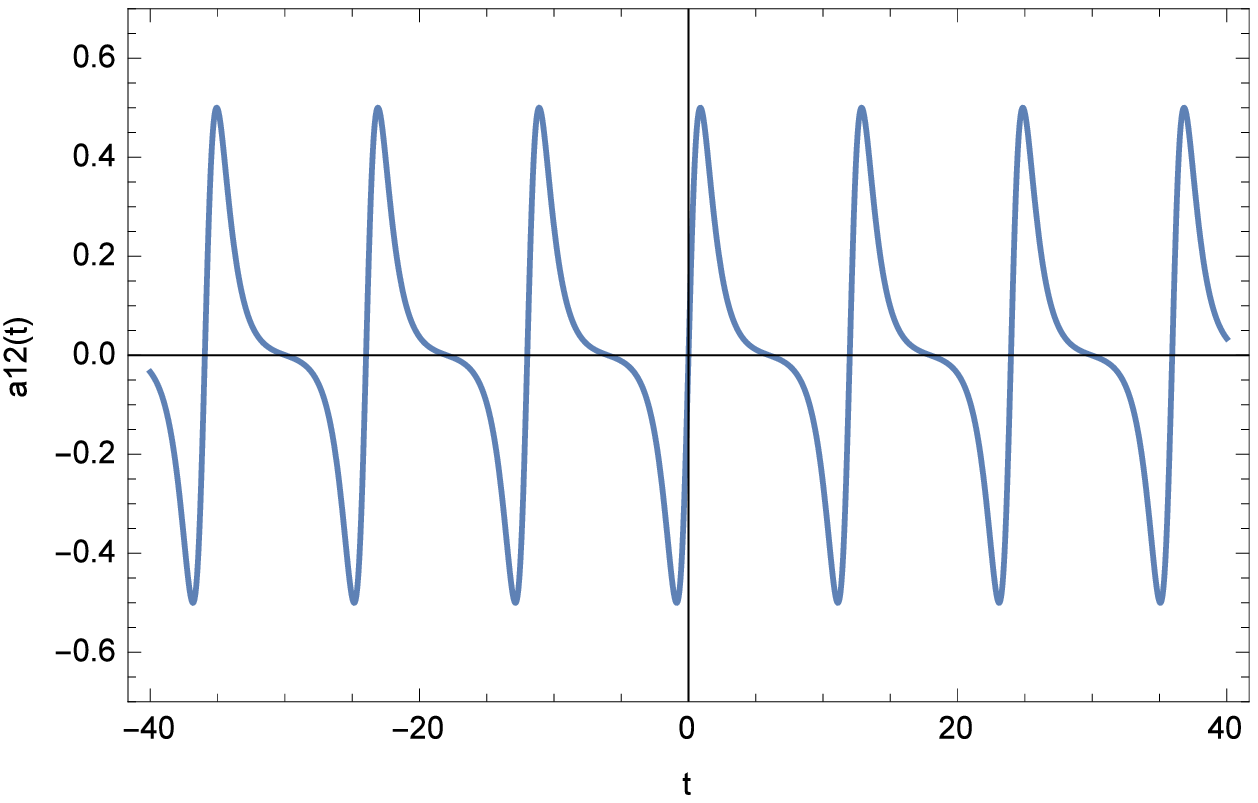}
\end{minipage}%
\begin{minipage}[c]{0.51\textwidth}
\includegraphics[width=2.5in,height=1.8in]{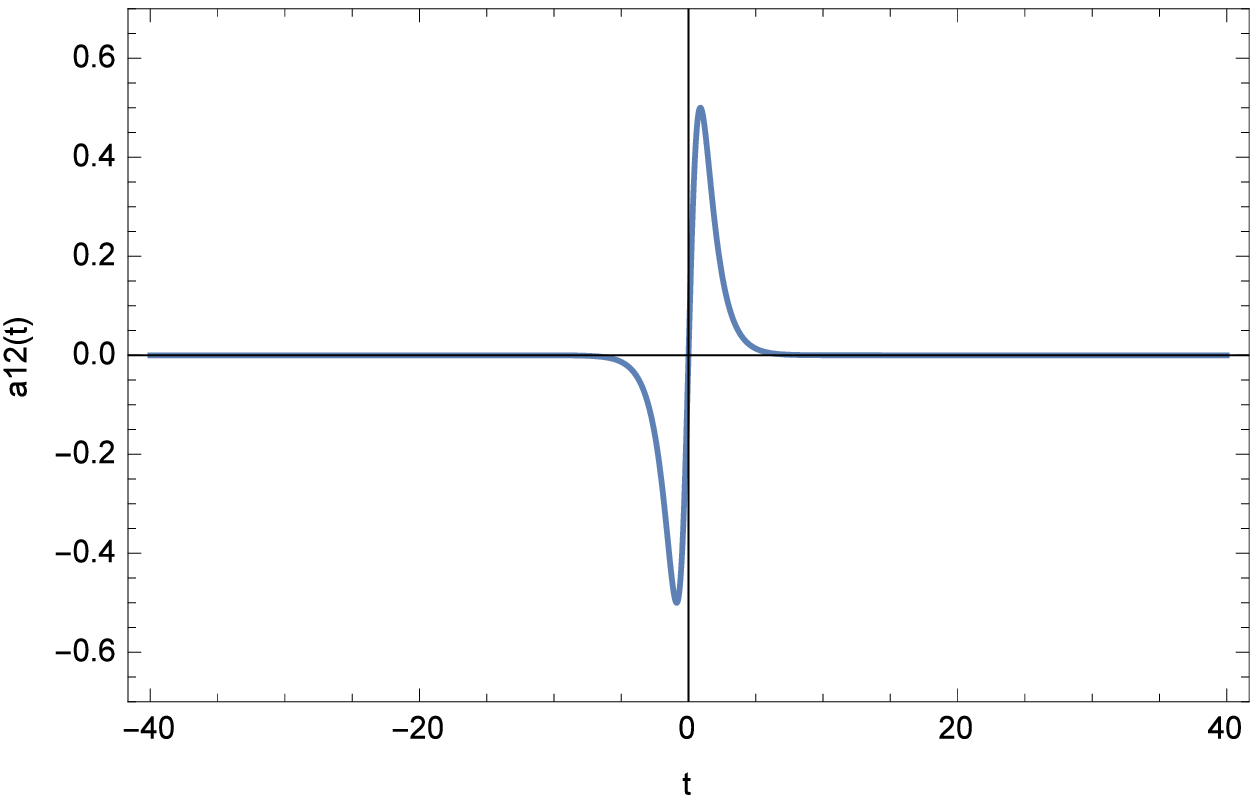}
\end{minipage}\\
\begin{minipage}[c]{0.51\textwidth}
\includegraphics[width=2.5in,height=1.8in]{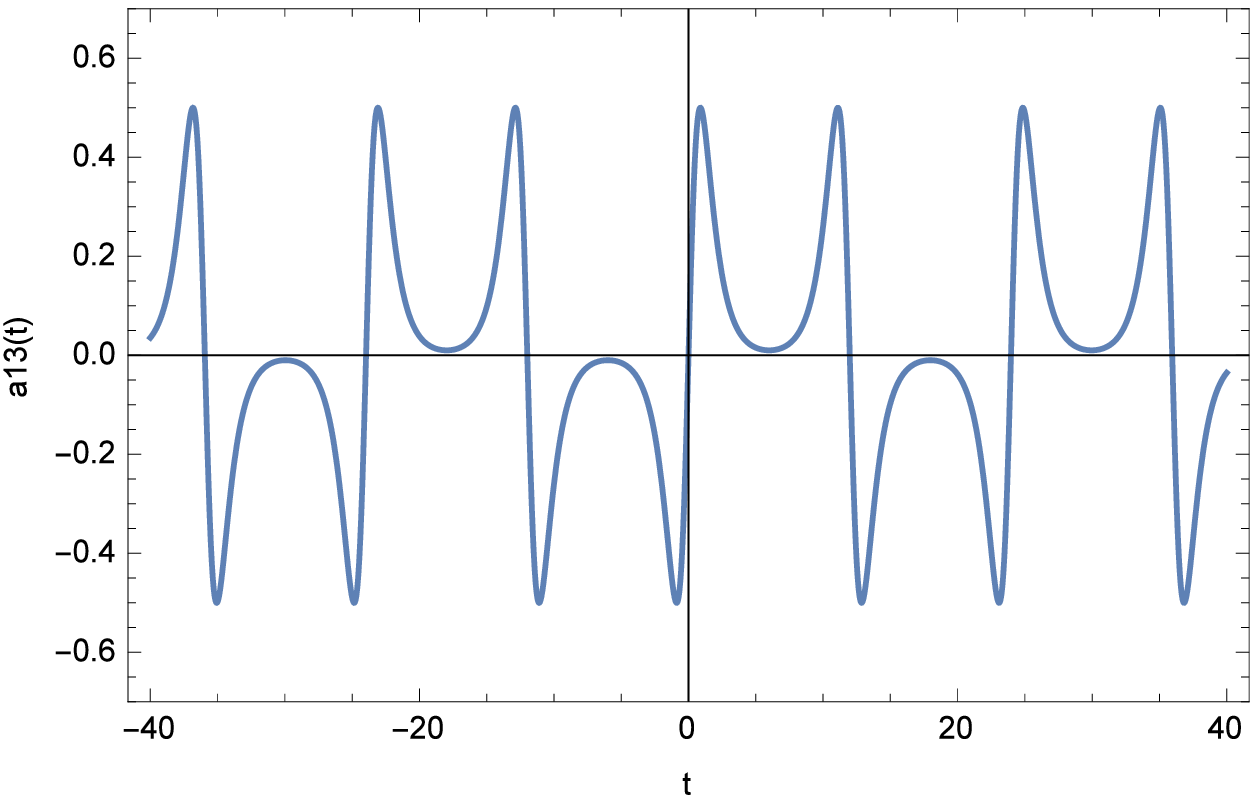}
\end{minipage}%
\begin{minipage}[c]{0.51\textwidth}
\includegraphics[width=2.5in,height=1.8in]{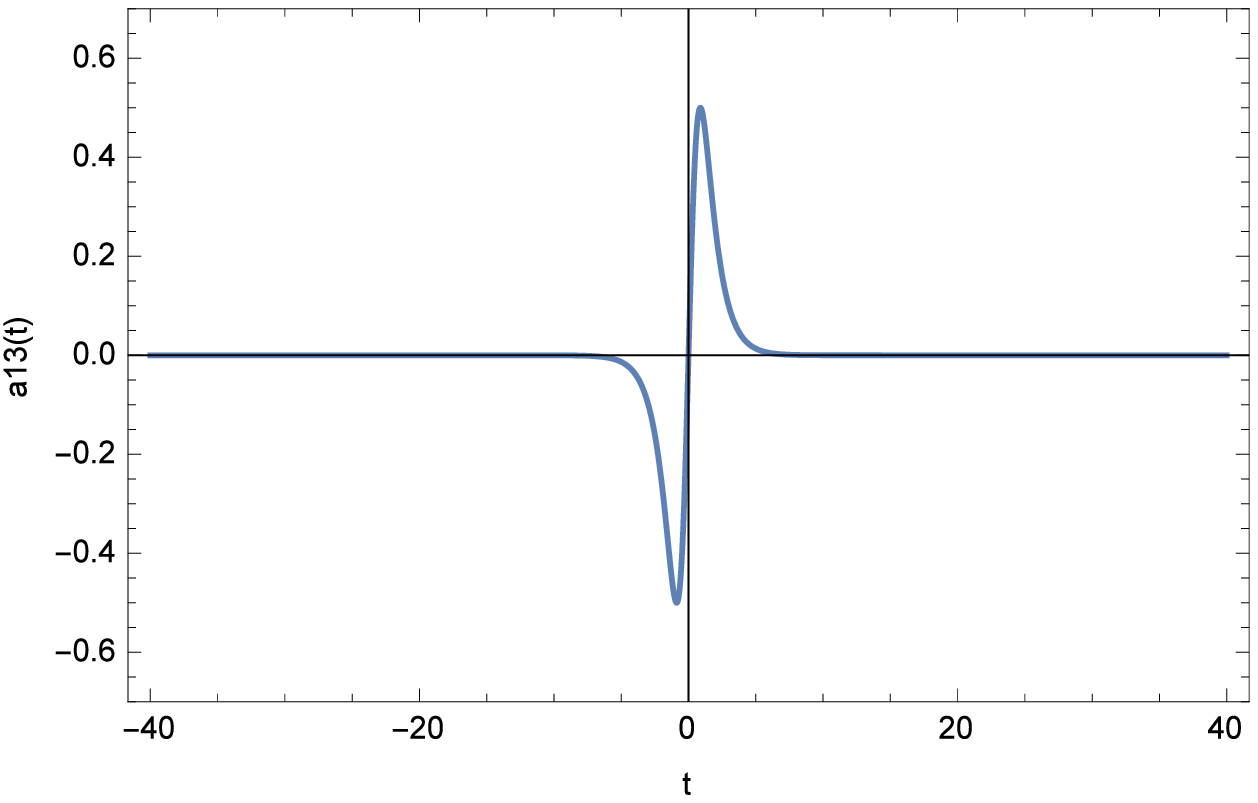}
\end{minipage}\\
\begin{minipage}[c]{0.51\textwidth}
\includegraphics[width=2.5in,height=1.8in]{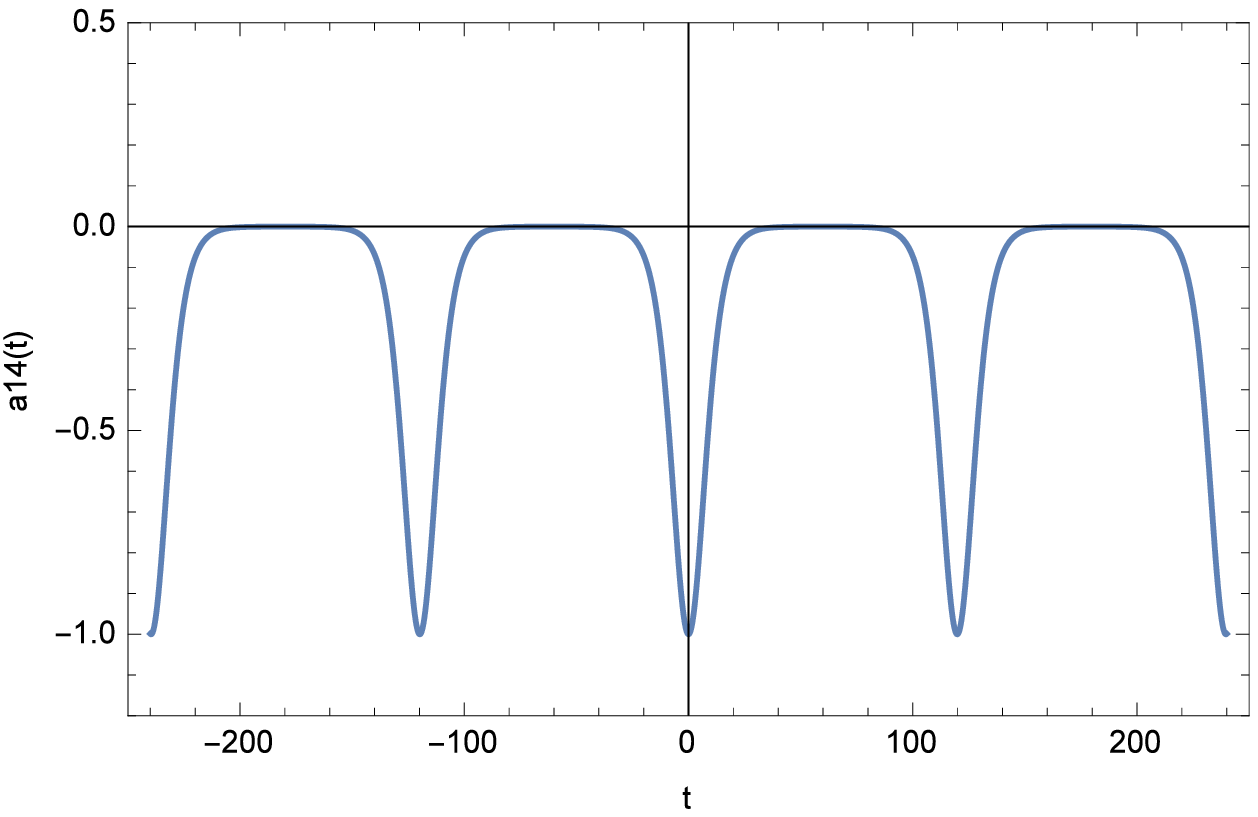}
\end{minipage}%
\begin{minipage}[c]{0.51\textwidth}
\includegraphics[width=2.5in,height=1.8in]{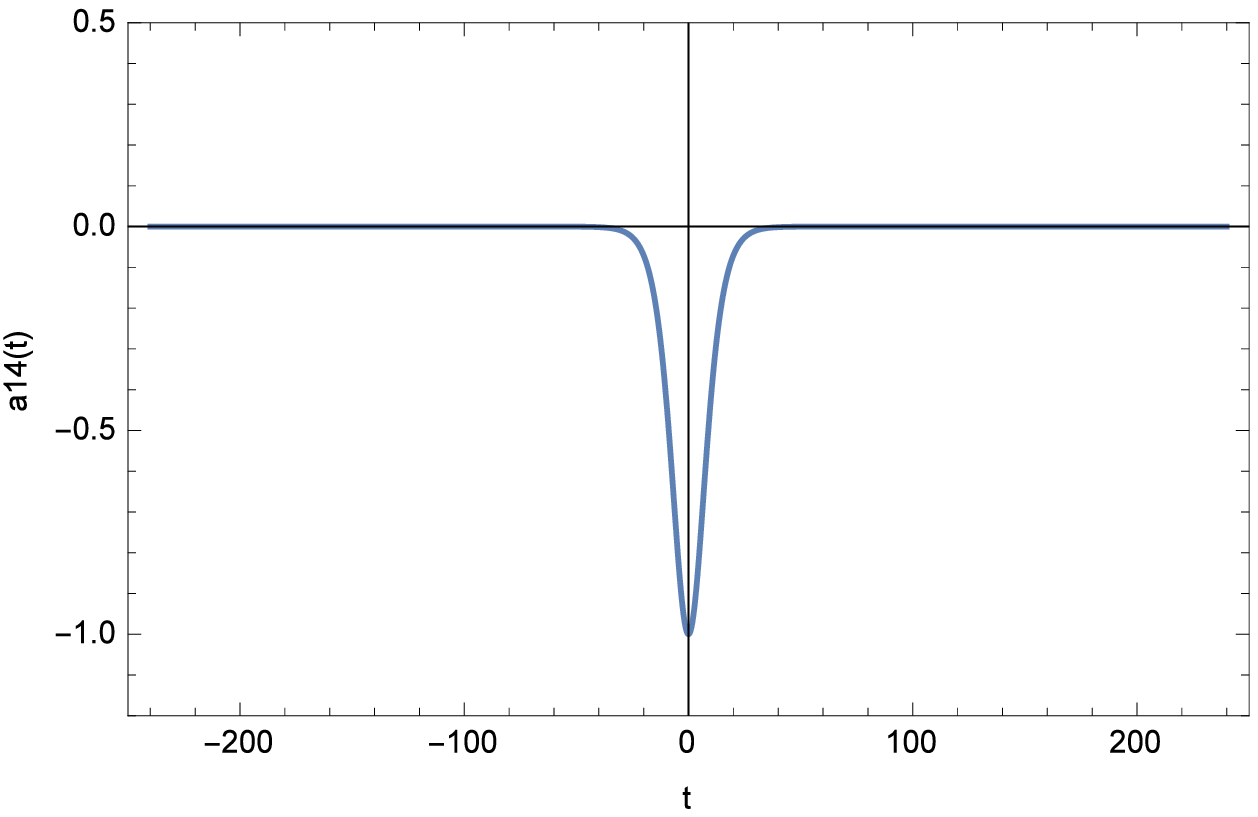}
\end{minipage}\\
\begin{minipage}[c]{0.51\textwidth}
\includegraphics[width=2.5in,height=1.8in]{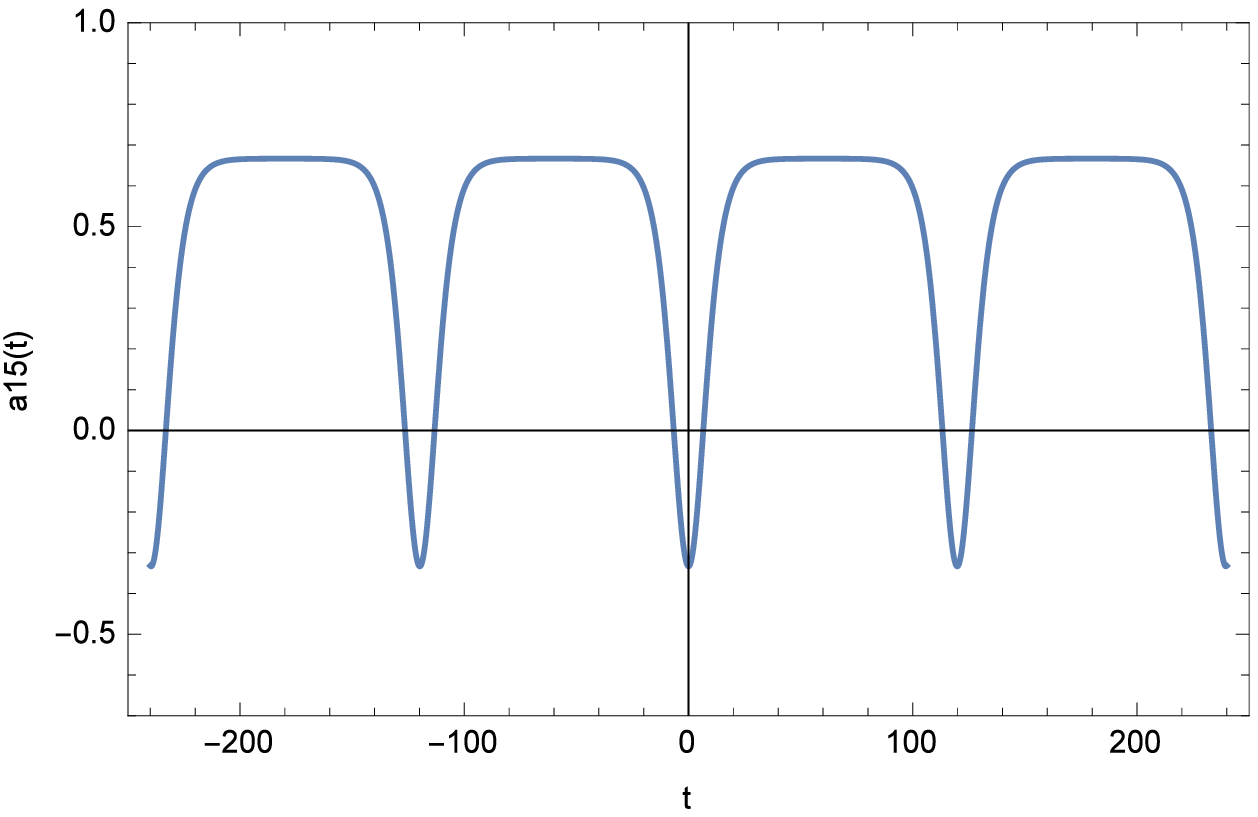}
\end{minipage}%
\begin{minipage}[c]{0.51\textwidth}
\includegraphics[width=2.5in,height=1.8in]{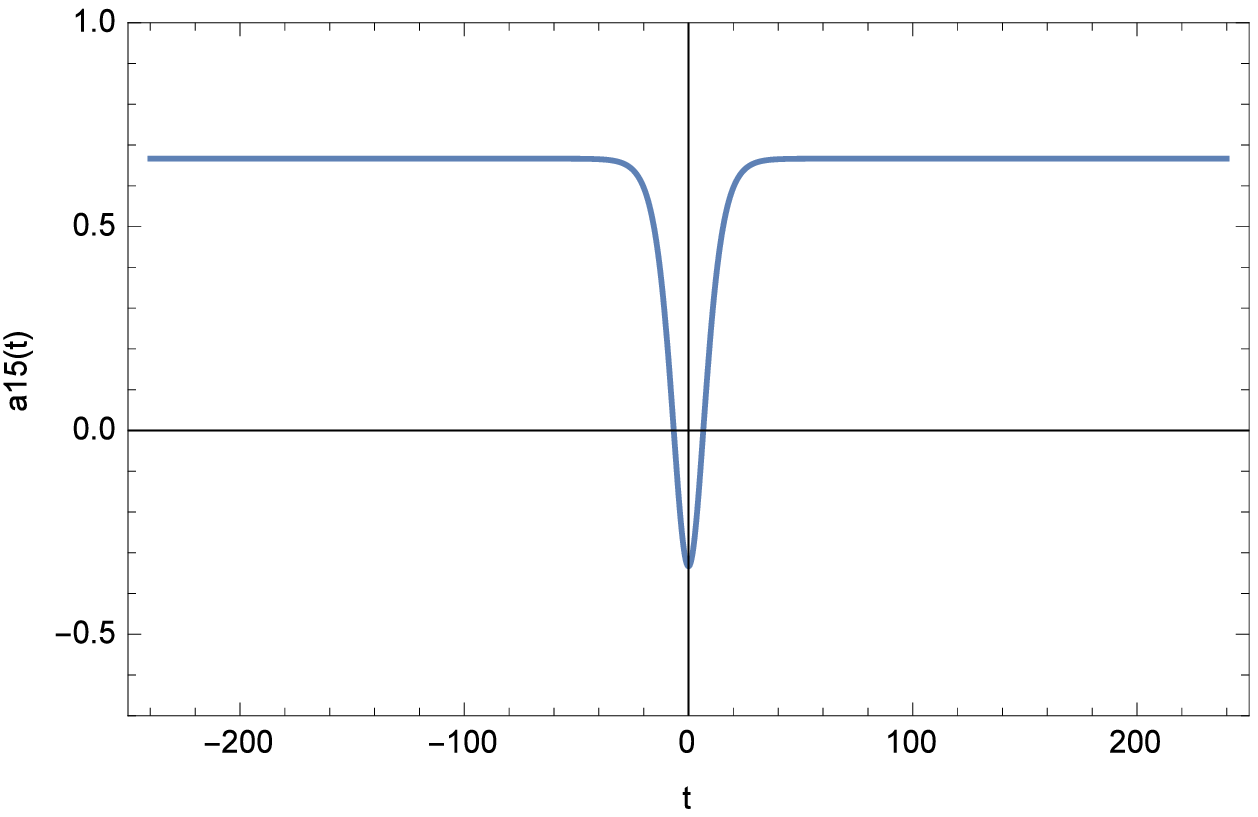}
\end{minipage}
\caption{Wave profiles of the five boundstates $a_{1j}$ ($j=1,2,3,4,5$) given by eqs. (\ref{eq:eig1})-(\ref{eq:eig4}), for $k= 0.98$ (left column) and $k= 1$ (right column).}
\label{fig:fandio5}
\end{figure*}

A not less relevant issue related to the study of nonlinear structures in optical media is their stability, once their existence is proven. This issue has been investigated at length for single pulse and dark soliton signals, particularly within the framework of linear stability analysis which provides an interesting way of testing their robustness against small-amplitude noises (see e.g. \cite{pelynov1}). For single-soliton signals the linear stability analysis of the wave equation (\ref{eq:one}) leads to an eignvalue problem for which the discrete spectrum consists of three localized modes with non-zero spatial modulations. From the standpoint of Physics these localized modes describe internal oscillations in the structure of the soliton signals that propagate together with the signals, and because of their non-zero energies they can be associated with "radiation-carrying" excitations in the propagating pulse background. \\
In the case of time-division multiplexed pulse states the interaction between pulses holds a key role in the stability of the structure as emphasized for instance in ref. \cite{bul}). To this last point, for the soliton-crystal structure obtained in the previous section the separation between individual pulses must be equal and therefore it is necessary to minimize their mutual interactions as well as the Gauss-Haus effect, two phenomena that tend to promote strong correlations between centers of mass of the interacting pulses. To explain the shape preservation (referred to as non-coalescence in ref. \cite{bul}) of individual pulses as well as their propagation in a fixed time slot within the soliton-crystal structure, Buryak et al. \cite{bul} suggested the possibility of oscillating tails on pulses that would compensate for the energy cost in pulse interactions with neighbour pulses through some effective potential. \\
In our investigations of stability of the soliton-crystal structure we shall proceed through the linear stability analysis, in follozing this approach we expect the emergence of additional internal modes, besides the boundstates proper to each individual pulse, that would be consistent with new internal degees of freedom due exclusively to the correlation of centers of mass of the interacting pulses. Proceeding with, consider a small disturbance $a_1(t)$ moving bound to the pulse-lattice envelope (\ref{eq:two}) of the soliton crystal, which we now denote $a_0(t)$. With this let 
\begin{equation}
a(t)= a_0(t) + a_1(t) \label{eq:exp}
\end{equation}
be the soliton-crystal envelope dressed with the noise $a_1(t)$, replacing $a(t)$ in the envelope equation derived from (\ref{eq:one}), to the linear order we find the following eigenvalue equation:
\begin{equation}
  \left[\frac{\partial^2}{\partial \tau^2} - 6k^2\,sn^2(\tau)\right] a_1(\tau)= E(k) a_1(\tau), \hskip 0.25truecm E(k)= 2\frac{3\beta_2\lambda^2 + \beta}{\beta_2\lambda^2}. \label{eq:eigen}
  \end{equation}
 This is Lam\'e's eigenvalue problem of second order \cite{dik1,dik2} in which $sn(\tau)$ is the Jacobi sine function of argument $\tau= \lambda\,t$. The eigenvalue problem (\ref{eq:eigen}) possesses both discrete and continuous modes all describing radiative modes of the soliton-crystal profile. Discrete modes are of particular interest since they represent localized excitations in the background of the soliton crystal. \\
 For eq. (\ref{eq:eigen}) there are five distinct localized modes \cite{dikan} namely: 
 \begin{equation}
 a_{11}(\tau)=a_{11}(k) cn(\tau) dn(\tau), \hskip 0.25truecm \beta_{11}= (5-k^2)\vert \beta_2\vert \lambda^2/2, \label{eq:eig1}
 \end{equation}
\begin{equation}
 a_{12}(\tau)=a_{12}(k) cn(\tau) sn(\tau), \hskip 0.25truecm \beta_{12}= (2-k^2)\vert \beta_2\vert \lambda^2/2, \label{eq:eig2}
 \end{equation}
 \begin{equation}
 a_{13}(\tau)=a_{13}(k) sn(\tau) dn(\tau), \hskip 0.25truecm \beta_{13}= (5-4k^2)\vert \beta_2\vert \lambda^2/2, \label{eq:eig3}
 \end{equation}
 
 \begin{eqnarray}
 a_{14, 15}(\tau)&=&a_{14}(k) \left[sn^2(\tau) - \frac{1+k^2}{3k^2} \mp \frac{\sqrt{1-k^2(1-k^2)}}{3k^2}\right], \nonumber \\ \beta_{14,15}&=& \left[2-k^2 \pm \frac{\sqrt{1-k^2(1-k^2)}}{2}\right] \vert \beta_2\vert \lambda^2, \label{eq:eig4}
 \end{eqnarray}
 where $a_{1j=1,2,3,4,5}(k)$ are their (constant) amplitudes. In fig. \ref{fig:fandio5} we sketched the five internal modes for $k=0.97$ (left column) and $k=1$ (right column), respectively. 

 Graphs in the right column of fig. \ref{fig:fandio5} are particularly instructive given that when $k=1$, the separation between pulses becomes infinite and the soliton crystal decays into a singl pulse. Thus, according to shape profiles in the right column of fig. \ref{fig:fandio5}, the single-pulse signal possesses three distinct internal modes including a single-pulse shaped, a symmetric two-pulse and an asymmetric two-pulse boundstates all with positive spatial modulations $\beta_{1j}$. Actually this feature of the internal-mode spectrum of the single pulse suggests that the two additional modes appearing when $k$ is smaller that one, as seen in the right column of fig. \ref{fig:fandio5}, are related to the correlation of centers of mass of the interacting pulses. \\
 As a final remark, we would like to stress that the linear stability analysis provides a very efficient way of probing the soliton stability and particularly its shape invariance under translation. To this last point, for the single-pulse signal one of the three localized modes is precisely the translation mode while the two others are related to localized excitations in the internal structure of the pulse interacting with the noise. We can determine which among the three boundstates is the translation mode by simply slighly translating the pulse by $\tau_0$ from some temporal position $t$, which yields: 
 \begin{equation}
  a_{1s}(\tau+\tau_0) \propto a_{1s}(\tau) + \tau_0 \, a'_{1s}(\tau) + 0(\tau_0^2), \label{eq:trans}
 \end{equation}
where $a'_{1s}(t)$ refers to time derivative of the field. For the single pulse the second term in the right hand side of formula (\ref{eq:trans}) is proportional to $- sech(\tau) tanh(\tau)$, this is actually the function formula (\ref{eq:eig2}) and the first graph in the right column of fig. \ref{fig:fandio5}. Similarly, an infinitesimal translation of the soliton crystal as a hole by $\tau_0$ leads to the same expression (\ref{eq:trans}) with a translation mode proportional to $sn(\tau) cn(\tau)$. However, if we translate individual pulses prior to temporal multiplexing we find an additional interaction-bearing excitation proportional to $sn(\tau) dn(\tau)$, coinciding with the second boundstate (\ref{eq:eig3}). Formulas (\ref{eq:eig2}) and  (\ref{eq:eig3}) show that the two modes tend to the same limit (same wave profile and same eigen energy) as $k\rightarrow 1$, so we can conclude that the correlation of centers of mass of individual pulses due to their interactions, broadens their internal-mode spectrum while inducing mode degeneracies. 
 
\section{\label{Sec:level5}Conclusion}
In summary we have established that in the regime of anomalous dispersion soliton crystal patterns can emerge in a nonlinear optical medium through a time division multiplexing process, and we proceeded with its reconstruction by considering an experimental setup consisting of periodically pumping identical pulses at high repetition rates. The Jacobi elliptic delta function was found to provide a good representation of the temporal envelope signal in the soliton-crystal state. The numerical reconstruction of the signal by time divsion multiplexing of a large but finite number of pulses, has led to a comb-like signal similar to those observed in passively mode-locked fiber lasers and confirmed that modulational instability of pulses fuels the formation of soliton crystals. At some threshold pumping limit a pulse train is formed and evolves into a more stable double-time scale signal as the pumping frequency increases. The self-preserving features of the pulse-soliton lattice offer a suitable means of long distance signal convoy in fiber waveguides, as widely illustrated in several past studies.\\
To end it would be instructive stressing that the issue of theoretical reconstruction of bound-soliton states observed experimentally in fiber laser media has already been addressed in few past works. Among these past studies the account given in ref. \cite{tang} on the formation of pulses and their subsequent evolution into muti-pulse soliton states, is consistent with our theoretical results. Indeed the authors observed that while the formation of fundamental pulses is governed mainly by the balance of nonlinearity and the anomalous group velocity dispersion, higher-energy pulses will split into identical lower-energy multisolitons with exactly the same physical properties. They found that the separation of neighboring pulses in the bound-soliton states was variational in the temporal domain, which suggests a time division multiplexing of identical pulses to form a periodic multi-soliton structure with temporal and spectral characteristics having a sizable dependence on the laser cavity roundtrip.

\begin{acknowledgments}
Jubgang Fandio thanks Atoneche Fred for enriching discussions on fiber lasers technology. A. M. Dikand\'e wishes to acknowledge supports of the Alexander von Humboldt foundationm and The World Academy of Science (TWAS). 
\end{acknowledgments}

\end{document}